\documentclass[11pt]{article}
\usepackage[superscript]{cite}
\usepackage{wordlike}
\usepackage{setspace}

\usepackage{hyperref}
\hypersetup{colorlinks,urlcolor=blue}
\usepackage{float}
\usepackage{pdflscape}
\usepackage{textcomp}
\usepackage{cite}
\usepackage{times}
\usepackage{latexsym}
\usepackage{hyperref}
\usepackage{booktabs}
\usepackage{amsmath}
\usepackage{threeparttable}
\usepackage{tabularx}
\usepackage{graphicx}
\usepackage{multirow}
\usepackage{longtable}
\usepackage{xcolor}
\usepackage{threeparttable}
\makeatletter
\definecolor{myblue1}{RGB}{47,84,150}
\newcommand{\printfnsymbol}[1]{%
  \textsuperscript{\@fnsymbol{#1}}%
}
\renewcommand{\section}{\@startsection%
  {section}%
  {0}%
  {0em}%
  {-\baselineskip}%
  {0.5\baselineskip}%
  {\color{myblue1}\Large\sffamily}}%
\renewcommand{\subsection}{\@startsection%
  {subsection}%
  {1}%
  {0em}%
  {-\baselineskip}%
  {0.5\baselineskip}%
  {\color{myblue1}\large\sffamily}}%
\makeatother

\renewcommand{\frame}{}

\title{\singlespacing Predicting risk of late age-related macular degeneration using deep learning}

\author{
Yifan Peng, PhD$^{1,\dagger}$, Tiarnan D. Keenan, BM BCh, PhD$^{2,\dagger}$, Qingyu Chen, PhD$^{1}$, Elvira Agr\'{o}n, MA$^{2}$, Alexis Allot, PhD$^1$, Wai T. Wong, MD$^{2}$, Emily Y. Chew, MD$^{2,*}$, Zhiyong Lu, PhD$^{1,*}$\\
1. National Center for Biotechnology Information (NCBI), National Library of Medicine (NLM), National Institutes of Health (NIH), Bethesda, Maryland, United States;\\
2. National Eye Institute (NEI), National Institutes of Health (NIH), Bethesda, Maryland, United States;\\ 
\textdagger{} These authors contributed equally to this work.\\
* To whom correspondence should be addressed: \url{echew@nei.nih.gov}; \url{zhiyong.lu@nih.gov}.\\
}

\date{}

\begin{document}

\maketitle

\section*{Abstract}

By 2040, age-related macular degeneration (AMD) will affect approximately 288 million people worldwide. Identifying individuals at high risk of progression to late AMD, the sight-threatening stage, is critical for clinical actions, including medical interventions and timely monitoring. Although deep learning has shown promise in diagnosing/screening AMD using color fundus photographs, it remains difficult to predict individuals’ risks of late AMD accurately. For both tasks, these initial deep learning attempts have remained largely unvalidated in independent cohorts. Here, we demonstrate how deep learning and survival analysis can predict the probability of progression to late AMD using 3,298 participants (over 80,000 images) from the Age-Related Eye Disease Studies AREDS and AREDS2, the largest longitudinal clinical trials in AMD. When validated against an independent test dataset of 601 participants, our model achieved high prognostic accuracy (five-year C-statistic 86.4 (95\% confidence interval 86.2-86.6)) that substantially exceeded that of retinal specialists using two existing clinical standards (81.3 (81.1-81.5) and 82.0 (81.8-82.3), respectively). Interestingly, our approach offers additional strengths over the existing clinical standards in AMD prognosis (e.g., risk ascertainment above 50\%) and is likely to be highly generalizable, given the breadth of training data from 82 US retinal specialty clinics. Indeed, during external validation through training on AREDS and testing on AREDS2 as an independent cohort, our model retained substantially higher prognostic accuracy than existing clinical standards. These results highlight the potential of deep learning systems to enhance clinical decision-making in AMD patients. 

\noindent\textbf{Keywords}

\noindent AMD, Deep Learning, Survival Analysis, Progression Prediction
\vspace{1em}

\section*{Introduction}

Age-related macular degeneration (AMD) is the leading cause of legal blindness in developed countries\cite{quartilho2016leading}. Through global demographic changes, the number of people with AMD worldwide is projected to reach 288 million by 2040\cite{wong2014global}. The disease is classified into early, intermediate, and late stages\cite{ferris2013clinical}. Late AMD, the stage associated with severe visual loss, occurs in two forms, geographic atrophy (GA) and neovascular AMD (NV). Making accurate time-based predictions of progression to late AMD is clinically critical. This would enable improved decision-making regarding: (i) medical treatments, especially oral supplements known to decrease progression risk, (ii) lifestyle interventions, particularly smoking cessation and dietary changes, and (iii) intensity of patient monitoring, e.g., frequent reimaging in clinic and/or tailored home monitoring programs\cite{areds2001report8,areds2013lutein,domalpally2019imaging,aredshomestudyresearchgroup2014randomized,guymer2019subthreshold}. It would also aid the design of future clinical trials, which could be enriched for participants with a high risk of progression events\cite{calaprice-whitty2019improving}.

Color fundus photography (CFP) is the most widespread and accessible retinal imaging modality used worldwide; it is the most highly validated imaging modality for AMD classification and prediction of progression to late disease\cite{ferris2005simplified,areds2005report17}. Currently, two existing standards are available clinically for using CFP to predict the risk of progression. However, both of these were developed using data from the AREDS only; now, an expanded dataset with more progression events is available following the completion of the AREDS2\cite{areds2013lutein}. Of the two existing standards, the most commonly used is the five-step Simplified Severity Scale (SSS)\cite{ferris2005simplified}. This is a points-based system whereby an examining physician scores the presence of two AMD features (macular drusen and pigmentary abnormalities) in both eyes of an individual. From the total score of 0-4, a five-year risk of late AMD is then estimated. The other standard is an online risk calculator\cite{klein2011risk}. Like the SSS, its inputs include the presence of macular drusen and pigmentary abnormalities; however, it can also receive the individual’s age, smoking status, and basic genotype information consisting of two SNPs (when available). Unlike the SSS system, the online risk calculator predicts the risk of progression to late AMD, GA, and NV at 1-10 years.

Both existing clinical standards face limitations. First, the ascertainment of the SSS features from CFP or clinical examination requires significant clinical expertise, typical in retinal specialists, but remains time-consuming and error-prone\cite{peng2018deepseenet}, even when performed by expert graders in a reading center\cite{areds2005report17}. Second, the SSS relies on two hand-crafted features and cannot receive other potentially risk-determining features. Recent work applying deep learning (DL)\cite{ching2018opportunities} has shown promise in the automated diagnosis and triage of conditions including cardiac, pediatric, dermatological, and retinal diseases\cite{peng2018deepseenet,grassmann2018deep,liang2019evaluation,hannun2019cardiologistlevel,gulshan2016development,poplin2018prediction,kermany2018identifying,esteva2017dermatologistlevel,ting2017development,ting2019deepa,raumviboonsuk2019deep,arcadu2019deep,abramoff2018pivotal}, but not in predicting the risk of AMD progression on a large scale or at the patient level\cite{burlina2018use}. Specifically, Burlina et al reported on the use of DL for estimating the AREDS 9-step severity grades of individual eyes, based on CFP in the AREDS dataset\cite{burlina2018use,yandex2015aggregating,schmidt-erfurth2018artificial}. However, this approach relied on previously published 5-year risk estimates at the severity class level\cite{areds2001report6}, rather than using the ground truth of actual progression/non-progression at the level of individual eyes, or the timing and subtype of any progression events. In addition, no external validation using an independent dataset was performed in that study. Babenko et al. proposed a model to predict risk of progression to neovascular AMD only (i.e. not late AMD or GA)\cite{yandex2015aggregating}. Importantly, the model was designed to predict 1-year risks only, i.e. at one fixed and very short interval only. Schmidt-Erfurth et al. proposed to use OCT images to predict progression to late AMD\cite{schmidt-erfurth2018artificial}. Specifically, they used a dataset (495 eyes, containing only 159 eyes that progressed to late AMD) with follow-up of 2 years. Their dataset was annotated by two retinal specialists (rather than unified grading by reading center experts using a published protocol).

\begin{figure}
	\centering
	\frame{\includegraphics[width=\textwidth,trim=0 3cm 1cm 3cm,clip]{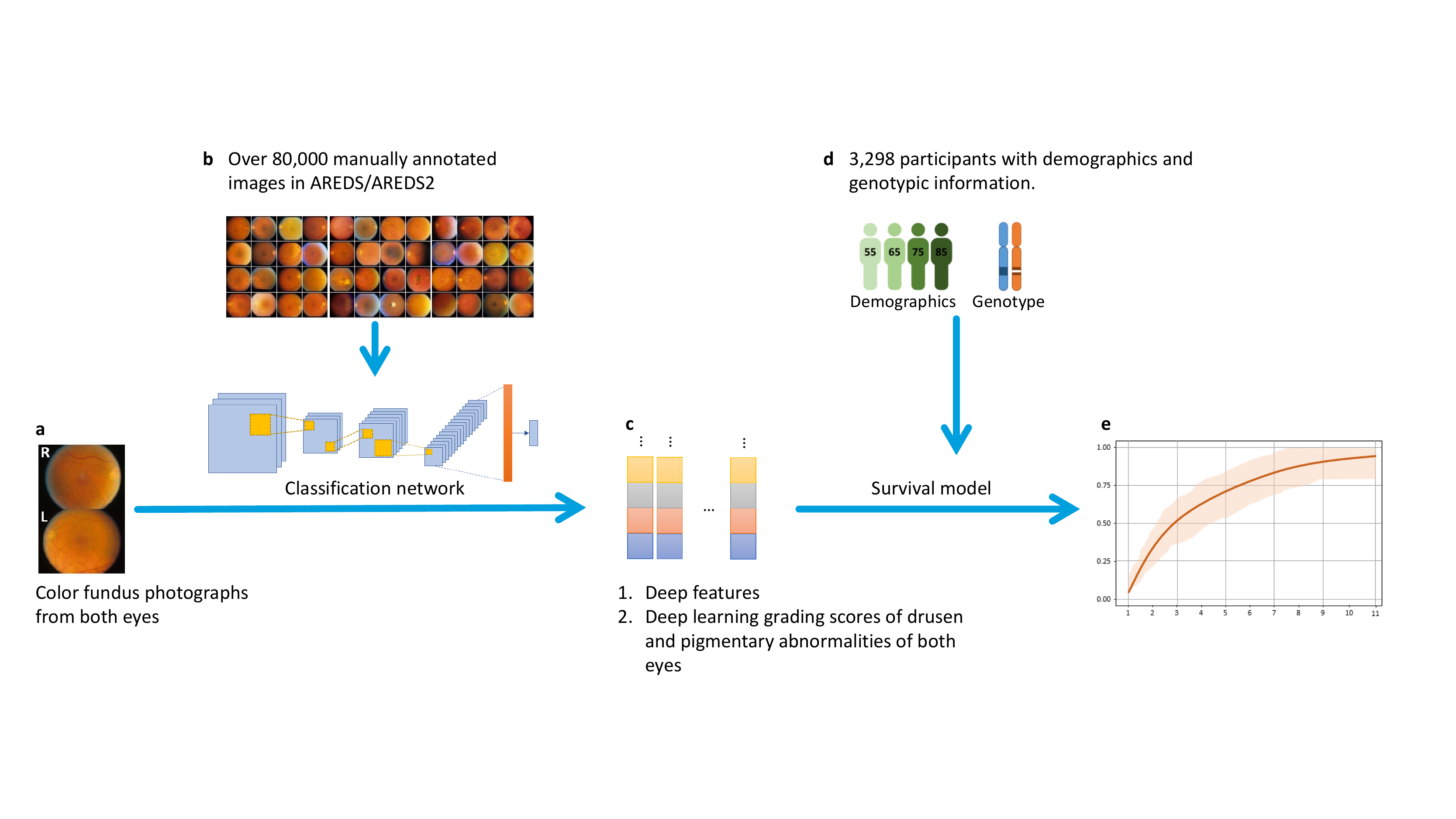}}
	\caption{\textbf{The two-step architecture of the framework}. (a) Raw color fundus photographs (CFP; field 2, i.e., 30\textdegree{} imaging field centered at the fovea). (b) Deep classification network, trained with CFP (all manually graded by reading center human experts). (c) Resulting deep features or deep learning grading. (d) Survival model, trained with imaging data and participant demographic information, with/without genotype information: ARMS2 rs10490924, CFH rs1061170, and 52-SNP Genetic Risk Score. (e) Late age-related macular degeneration survival probability.}
	\label{fig:architecture}
\end{figure}

\begin{figure}
	\centering
	\frame{\includegraphics[width=.8\textwidth,trim=0 0 4cm 0,clip]{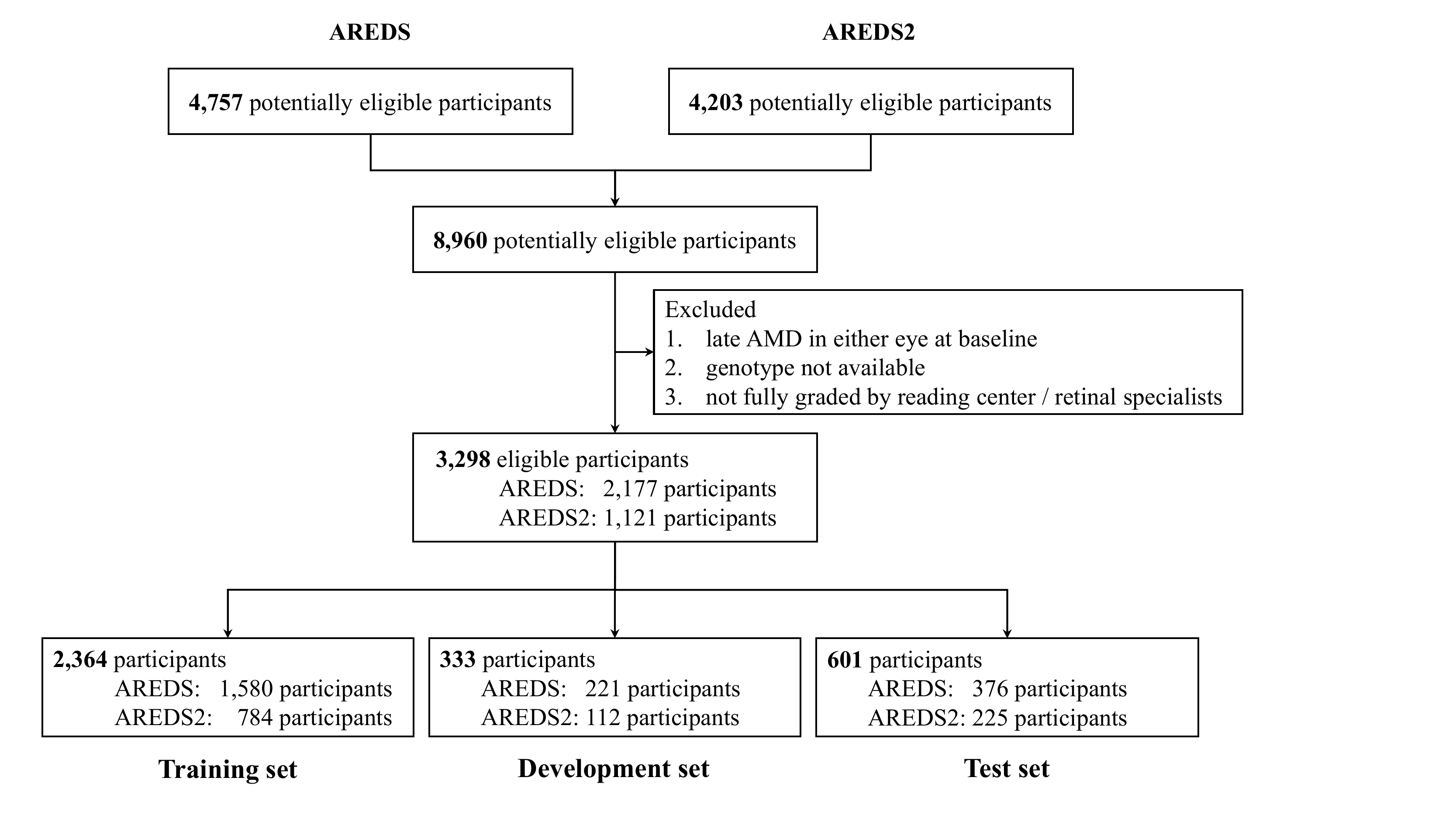}}
	\caption{\textbf{Creation of the study data sets}. To avoid ‘cross-contamination’ between the training and test datasets, no participant was in more than one group.}
	\label{fig:data}
\end{figure}

Here, we developed a DL architecture to predict progression with improved accuracy and transparency in two steps: image classification followed by survival analysis (Figure~\ref{fig:architecture}). The model is developed and clinically validated on two datasets from the Age-Related Eye Disease Studies (AREDS\cite{areds1999report1} and AREDS2\cite{areds22012report1}), the largest longitudinal clinical trials in AMD (Figure~\ref{fig:data}). The framework and datasets are described in detail in the Methods section.
\vspace{1em}

Our framework has several important strengths. First, it performs progression predictions directly from CFP over a wide time interval (1-12 years). Second, training and testing were based on the ground truth of reading center-graded progression events at the level of individuals. Both training and testing benefitted from an expanded dataset with many more progression events, achieved by using data from the AREDS2 alongside AREDS, for the first time in DL studies. Third, our framework can predict the risk not only of late AMD, but also of GA and NV separately. This is important since treatment approaches for the two subtypes of late AMD are very different: NV needs to be diagnosed extremely promptly, since delay in access to intravitreal anti-VEGF injections is usually associated with very poor visual outcomes\cite{flaxel2020agerelateda}, while various therapeutic options to slow GA enlargement are under investigation\cite{liao2020complement,fleckenstein2018progression}. Finally, the two-step approach has important advantages. By separating the DL extraction of retinal features from the survival analysis, the final predictions are more explainable and biologically plausible, and error analysis is possible. By contrast, end-to-end ‘black-box’ DL approaches are less transparent and may be more susceptible to failure\cite{ting2019deep}.

\section*{Results}

\subsection*{Deep learning models trained on the combined AREDS/AREDS2 training sets and validated on the combined AREDS/AREDS2 test sets}

The characteristics of the participants are shown in Table~\ref{tab:characteristics}. The characteristics of the images are shown in Supplementary Table~\ref{tab:number}. The overall framework of our method is shown in Figure~\ref{fig:architecture} and described in detail in the Methods section. In short, first, a deep convolutional neural network (CNN) was adapted to (i) extract multiple highly discriminative deep features, or (ii) estimate grades for drusen and pigmentary abnormalities (Figure~\ref{fig:architecture} b to c). Second, a Cox proportional hazards model was used to predict probability of progression to late AMD (and GA/NV, separately), based on the deep features (‘deep features/survival’) or the DL grading (‘DL grading/survival’) (Figure~\ref{fig:architecture} d and e). In this step, additional participant information could be added, such as age, smoking status, and genetics. Separately, all of the baseline images in the test set were graded by 88 (AREDS) and 192 (AREDS2) retinal specialists. By using these grades as input to either the SSS or the online calculator, we computed the prediction results of the two existing standards: ‘retinal specialists/SSS’ and ‘retinal specialists/calculator’.

\begin{table}
\caption{Characteristics of AREDS and AREDS2 participants. The AREDS contained a wide spectrum of baseline disease severity, from no age-related macular degeneration (AMD) to high-risk intermediate AMD. The AREDS2 contained a high level of baseline disease severity, i.e., high proportion of eyes at high risk of progression to late AMD.}
\label{tab:characteristics}
\begin{center}
\footnotesize
\begin{tabularx}{\textwidth}{Xll}
	\toprule
	Characteristicss & AREDSs & AREDS2\\
	\midrule
	\multicolumn{3}{l}{Participants characteristics}\\	
	\hspace{1em}Number of participants & 2,177 & 1,121\\
	\hspace{1em}Age, mean (SD), y & 68.4 (4.8) & 70.9 (7.9)\\
	\hspace{1em}Smoking history (never, former, current), \% & 49.6/45.6/4.8 & 45.9/48.5/5.5\\
	\hspace{1em}CFH rs1061170 (TT/CT/CC), \% & 33.3/46.4/20.3 & 17.8/41.7/40.5\\
	\hspace{1em}ARMS2 rs10490924 (GG/GT/TT), \% & 55.2/36.5/8.3 & 40.5/42.8/16.7\\
	\hspace{1em}AMD Genetic Risk Score, mean (SD) & 14.2 (1.4) & 15.2 (1.3)\\
	\hspace{1em}Follow-up, Median (IQR), y & 10.0 (3.0) & 5.0 (2.0)\\
	\multicolumn{3}{l}{Progression to late AMD (classified by Reading Center)}\\	
	\hspace{1em}Late AMD: \% of participants at year 1/2/3/4/5/all years & 1.5/3.9/6.68/8.6/10.7/18.5 & 9.1/16.2/23.8/32.6/38.1/38.8\\
	\hspace{1em}GA: \% of participants at year 1/2/3/4/5/all years & 0.6/1.5/2.8/4.0/5.1/10.1 & 4.8/9.0/13.0/17.8/20.8/21.0\\
	\hspace{1em}NV: \% of participants at year 1/2/3/4/5/all years & 0.9/2.4/4.0/4.6/5.5/8.4 & 4.3/7.2/10.8/14.7/17.3/17.8\\
	\bottomrule
\end{tabularx}
\end{center}
\small
\noindent SD - standard deviation. y - year. IQR - interquartile range. AMD - Age-related Macular Degeneration.
\end{table}

The prediction accuracy of the approaches was compared using the five-year C-statistic as the primary outcome measure. The five-year C-statistic of the two DL approaches met and substantially exceeded that of both existing standards (Table~\ref{tab:cindex}). For predictions of progression to late AMD, the five-year C-statistic was 86.4 (95\% confidence interval 86.2-86.6) for deep features/survival, 85.1 (85.0-85.3) for DL grading/survival, 82.0 (81.8-82.3) for retinal specialists/calculator, and 81.3 (81.1-81.5) for retinal specialists/SSS. For predictions of progression to GA, the equivalent results were 89.6 (89.4-89.8), 87.8 (87.6-88.0), and 82.6 (82.3-82.9), respectively; these are not available for retinal specialists/SSS, since the SSS does not make separate predictions for GA or NV. For predictions of progression to NV, the equivalent results were 81.1 (80.8-81.4), 80.2 (79.8-80.5), and 80.0 (79.7-80.4), respectively.

\begin{table}
\caption{The C-statistic (95\% confidence interval) of the survival models in predicting risk of progression to late age-related macular degeneration on the combined AREDS/AREDS2 test sets (601 participants).\label{tab:cindex}}
\begin{center}
\scriptsize
\begin{tabularx}{\textwidth}{X@{~~}cccccc}
\toprule
Models & 1 & 2 & 3 & 4 & 5 & All years\\
\midrule
Late AMD &  &  &  &  &  & \\
\hspace{1em}Deep features/survival  & 87.8(87.5,88.1) & 85.8(85.4,86.2) & 86.3(86.1,86.6) & 86.7(86.5,86.9) & 86.4(86.2,86.6) & 86.7(86.5,86.8)\\
\hspace{1em}DL grading/survival  & 84.9(84.6,85.3) & 84.1(83.8,84.4) & 84.8(84.5,85.0) & 84.8(84.6,85.0) & 85.1(85.0,85.3) & 84.9(84.6,85.3)\\
\hspace{1em}Retinal specialists/calculator & - & 78.3(77.9,78.8) & 81.8(81.5,82.1) & 82.7(82.4,82.9) & 82.0(81.8,82.3) & -\\
\hspace{1em}Retinal specialists/SSS & - & - & - & - & 81.3(81.1,81.5) & -\\
Geographic atrophy &  &  &  &  &  & \\
\hspace{1em}Deep features/survival  & 89.2(88.9,89.6) & 91.0(90.7,91.2) & 88.7(88.4,88.9) & 89.1(88.9,89.3) & 89.6(89.4,89.8) & 89.2(89.1,89.4)\\
\hspace{1em}DL grading/survival  & 88.6(88.3,88.9) & 86.6(86.4,86.9) & 87.6(87.4,87.9) & 88.1(87.9,88.3) & 87.8(87.6,88.0) & 88.6(88.3,88.9)\\
\hspace{1em}Retinal specialists/calculator & - & 77.5(76.9,78.0) & 81.2(80.8,81.6) & 82.0(81.6,82.3) & 82.6(82.3,82.9) & -\\
\hspace{1em}Retinal specialists/SSS & - & - & - & - & - & -\\
Neovascular AMD &  &  &  &  &  & \\
\hspace{1em}Deep features/survival  & 85.4(84.9,85.9) & 77.9(77.3,78.5) & 81.7(81.3,82.1) & 81.7(81.4,82.1) & 81.1(80.8,81.4) & 82.1(81.8,82.4)\\
\hspace{1em}DL grading/survival  & 78.0(77.4,78.6) & 78.4(77.9,78.9) & 79.2(78.8,79.6) & 79.0(78.7,79.4) & 80.2(79.8,80.5) & 78.0(77.4,78.6)\\
\hspace{1em}Retinal specialists/calculator  & - & 75.5(74.8,76.2) & 81.7(81.2,82.1) & 81.4(81.0,81.8) & 80.0(79.7,80.4) & -\\
\hspace{1em}Retinal specialists/SSS1 & - & - & - & - & - & -\\
\bottomrule
\end{tabularx}
\end{center}
\small
\noindent Retinal specialists/SSS – makes predictions at one fixed interval of five years and for late AMD only (i.e., not by disease subtype); unlike all other models, for SSS, late AMD is defined as NV or central GA (instead of NV or any GA); Please refer to the Supplementary Table~\ref{tab:multivariate} for results using genotype information, and Supplementary Table~\ref{tab:cindexcombine} for multivariate analysis. AMD - age-related macular degeneration; DL - deep learning; SSS - Simplified Severity Scale.
\end{table}

Similarly, for predictions at 1-4 years, the C-statistic was higher in all cases for the two DL approaches than the retinal specialists/calculator approach. Of the two DL approaches, the C-statistics of deep features/survival were higher in most cases than those of DL grading/survival. Predictions at these time intervals were not available for retinal specialists/SSS, since the SSS does not make predictions at any interval other than five years. Regarding the separate predictions of progression to GA and NV, deep features/survival also provided the most accurate predictions at most time intervals. Overall, DL-based image analysis provided more accurate predictions than those from retinal specialist grading using the two existing standards. For deep feature extraction, this may reflect the fact that DL is unconstrained by current medical knowledge and not limited to two hand-crafted features.

\begin{figure}
	\centering
	\frame{\includegraphics[width=\textwidth,trim=0 3cm 0 4cm,clip]{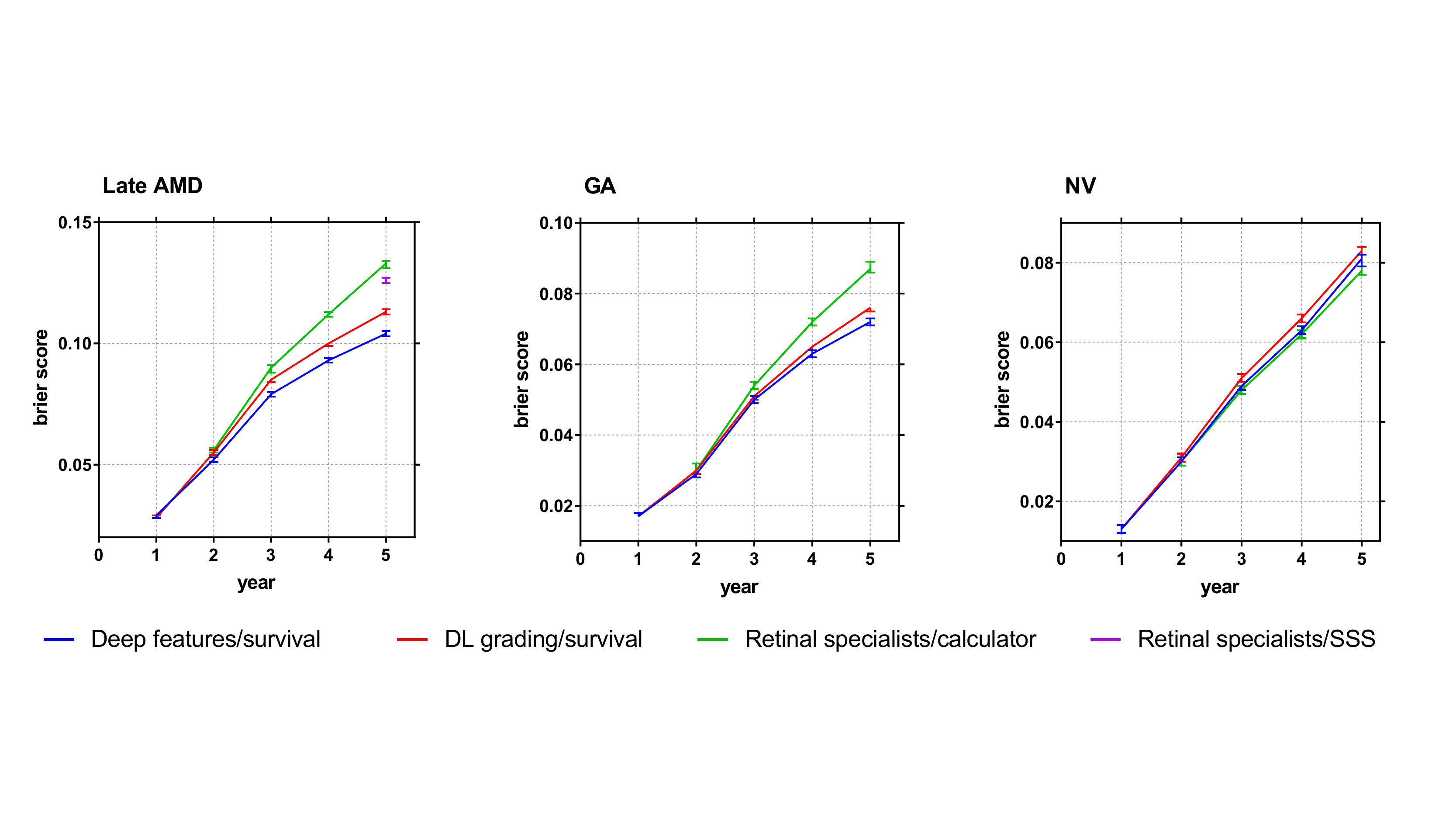}}
	\caption{\textbf{Prediction error curves}. Prediction error curves of the survival models in predicting risk of progression to late age-related macular degeneration on the combined AREDS/AREDS2 test sets (601 participants), using the Brier score (95\% confidence interval). }
	\label{fig:curves}
\end{figure}

In addition, the prediction calibrations were compared using the Brier score (Figure~\ref{fig:curves}). For five-year predictions of late AMD, the Brier score was lowest (i.e., optimal) for deep features/survival. We also split the data into 5 groups based on the AREDS SSS at baseline. We compared calibration plots for deep features/survival, DL grading/survival, and retinal specialist/survival with the actual progression data for the 5 groups (Supplement Figure~\ref{fig:survival}). The actual progression data for the five groups are shown in lines (Kaplan–Meier curves) and the predictions of our models are shown in lines with markers. The figure shows that the predictions of the deep features/survival model correspond better to the actual progression data than those of the other two models.

\subsection*{Deep learning models trained separately on individual cohorts (either AREDS or AREDS2) and validated on the combined AREDS/AREDS2 test sets}

Models trained on the combined AREDS/AREDS2 cohort (Table~\ref{tab:cindex}) were substantially more accurate than those trained on either individual cohort (Table~\ref{tab:cindex5}), with the additional advantage of improved generalizability. Indeed, one challenge of DL has been that generalizability to populations outside the training set can be variable. In this instance, the widely distributed sites and diverse clinical settings of AREDS/AREDS2 participants, together with the variety of CFP cameras used, help provide some assurance of broader generalizability.

\begin{table}
\centering
\caption{The 5-year C-statistic (95\% CI) results of models trained on only AREDS or only AREDS2, and validated on the combined AREDS/AREDS2 test sets (601 participants), without using genotype information.}
\label{tab:cindex5}
\begin{tabular}{lcc}
\toprule	
Models & Trained on AREDS & Trained on AREDS2\\
\midrule
Late AMD &  & \\
\hspace{1em}Deep features/survival & 85.7 (85.5, 85.9) & 83.9 (83.7, 84.1)\\
\hspace{1em}DL grading/survival    & 84.7 (84.5, 84.9) & 82.1 (81.8, 82.3)\\
Geographic atrophy &  & \\
\hspace{1em}Deep features/survival & 89.3 (89.1, 89.5) & 84.7 (84.4, 85.0)\\
\hspace{1em}DL grading/survival    & 90.2 (90.0, 90.4) & 85.2 (84.9, 85.5)\\
Neovascular AMD &  & \\
\hspace{1em}Deep features/survival & 79.6 (79.3, 80.0) & 74.0 (73.6, 74.5)\\
\hspace{1em}DL grading/survival    & 76.6 (76.2, 76.9) & 75.5 (75.1, 75.9)\\
\bottomrule
\end{tabular}
\end{table}

\subsection*{Deep learning models trained on AREDS and externally validated on AREDS2 as an independent cohort}

In separate experiments, to externally validate the models on an independent dataset, we trained the models on AREDS (2,177 participants) and tested them on AREDS2 (1,121 participants). Table~\ref{tab:cindex5genotype} shows that deep features/survival demonstrated the highest accuracy of five-year predictions in all scenarios, and DL grading/survival also had higher accuracy than retinal specialists/calculator.

\begin{table}
	\caption{The 5-year C-statistic (95\% CI) results of models trained on the entire AREDS and tested on the entire AREDS2 (1,121 participants), without using genotype information.}
	\label{tab:cindex5genotype}
\begin{center}
	\begin{tabular}{lc}
\toprule		
Models & Tested on the entire AREDS2\\
\midrule
Late AMD & \\
\hspace{1em}Deep features/survival 		   	& 71.0 (70.2, 71.7)\\
\hspace{1em}DL grading/survival  		   	& 69.7 (68.9, 70.5)\\
\hspace{1em}Retinal specialists/calculator 	& 63.9 (63.2, 64.6)\\
\hspace{1em}Retinal specialists/SSS1 	   	& 62.5 (62.3, 62.7)\\
Geographic atrophy & \\
\hspace{1em}Deep features/survival         	& 75.3 (74.5, 76.0)\\
\hspace{1em}DL grading/survival         	& 75.0 (74.0, 76.0)\\
\hspace{1em}Retinal specialists/calculator 	& 64.4 (63.6, 65.2)\\
\hspace{1em}Retinal specialists/SSS       	& -\\
Neovascular AMD & \\
\hspace{1em}Deep features/survival 			& 62.8 (61.9, 63.8)\\
\hspace{1em}DL grading/survival  			& 61.8 (61.0, 62.7)\\
\hspace{1em}Retinal specialists/calculator 	& 61.8 (60.8, 62.9)\\
\hspace{1em}Retinal specialists/SSS1 		& -\\
\bottomrule
	\end{tabular}
\end{center}
 Retinal specialists/SSS – makes predictions at one fixed interval of five years and for late AMD only (i.e., not by disease subtype); unlike all other models, for SSS, late AMD is defined as NV or central GA (instead of NV or any GA);
\end{table}

\subsection*{Survival models with additional input of genotype}

For all approaches possible, the predictions were tested with the additional input of genotype (Table~\ref{tab:accuracy}). Interestingly, adding the genotype data, even the 52 SNP-based Genetic Risk Score (GRS; see Methods) available only in rare research contexts\cite{fritsche2016large}, did not improve the accuracy for deep features/survival or DL grading/survival; by contrast, adding just two SNPs (the maximum handled by the calculator) did improve the accuracy modestly for the retinal specialists/calculator approach. Multivariate analysis (Supplementary Table~\ref{tab:multivariate}) demonstrated that deep features/DL grading, age, and AMD GRS contributed significantly to the survival models. The non-reliance of the DL approaches on genotype information favors their accessibility, as genotype data are typically unavailable for patients currently seen in clinical practice. It suggests that adding genotype information may partially compensate for the inferior accuracy obtained from human gradings, but contributes little to the accuracy of DL approaches, particularly deep feature extraction.

\begin{table}
\scriptsize
\caption{The accuracy (C-statistic, 95\% confidence interval) of the three different approaches in predicting risk of progression to late AMD on the combined AREDS/AREDS2 test sets (601 participants), with the inclusion of accompanying genotype information. a, Use of CFH rs1061170 and ARMS2 rs10490924 status only. b, use of CFH/ARMS2 and the 52 SNP-based AMD Genetic Risk Score. The AMD online calculator is able to receive only CFH rs1061170 and ARMS2 rs10490924 status, not the 52 SNP-based AMD Genetic Risk Score.}
	\label{tab:accuracy}
\vspace{1em}
{\normalsize \textbf{a}}
\begin{center}
\begin{tabular}{lc@{~~}ccccc}
\toprule		
Models & 1 & 2 & 3 & 4 & 5 & All years\\
\midrule
Late AMD &  &  &  &  &  & \\
\hspace{1em}Deep features/survival  & 88.8(88.0,89.7) & 85.4(84.4,86.4) & 86.1(85.2,86.9) & 86.9(86.1,87.6) & 86.6(85.8,87.4) & 86.8(86.1,87.6)\\
\hspace{1em}DL grading/survival  & 89.4(88.2,90.6) & 83.8(82.4,85.1) & 84.0(83.0,85.1) & 85.0(84.2,85.8) & 84.9(84.2,85.6) & 85.5(84.9,86.2)\\
\hspace{1em}Retinal specialists/calculator & - & 78.7(78.3,79.1) & 82.4(82.1,82.7) & 83.6(83.3,83.8) & 83.1(82.8,83.3) & -\\
Geographic atrophy &  &  &  &  &  & \\
\hspace{1em}Deep features/survival  & 89.5(88.4,90.5) & 90.7(89.9,91.5) & 88.6(87.3,89.8) & 89.0(88.0,89.9) & 89.5(88.7,90.3) & 89.4(88.7,90.2)\\
\hspace{1em}DL grading/survival  & 87.0(86.0,88.0) & 86.5(85.6,87.5) & 85.9(85.0,86.8) & 87.0(86.3,87.7) & 87.6(87.0,88.2) & 87.5(87.0,88.0)\\
\hspace{1em}Retinal specialists/calculator & - & 77.6(77.1,78.1) & 81.8(81.5,82.2) & 83.0(82.7,83.3) & 83.2(83.0,83.5) & -\\
Neovascular AMD &  &  &  &  &  &\\ 
\hspace{1em}Deep features/survival  & 85.8(84.6,87.0) & 79.1(77.0,81.2) & 81.9(80.6,83.3) & 82.4(81.2,83.6) & 81.7(80.7,82.7) & 82.6(81.7,83.5)\\
\hspace{1em}DL grading/survival  & 87.3(86.3,88.3) & 77.3(75.1,79.5) & 77.3(75.6,79.0) & 78.7(77.1,80.2) & 78.8(77.4,80.1) & 80.0(78.8,81.3)\\
\hspace{1em}Retinal specialists/calculator  & - & 77.3(76.6,77.9) & 82.0(81.5,82.4) & 82.4(82.1,82.8) & 80.9(80.6,81.3) & -\\
\bottomrule
\end{tabular}
\end{center}
{\normalsize \textbf{b}}
\begin{center}
\begin{tabular}{lcccccc}
\toprule
Models & 1 & 2 & 3 & 4 & 5 & All years\\
\midrule
Late AMD &  &  &  &  &  & \\
\hspace{1em}Deep features/survival  & 87.9(87.6,88.2) & 84.8(84.4,85.2) & 85.9(85.6,86.2) & 86.2(86.0,86.4) & 86.0(85.8,86.2) & 86.4(86.2,86.6)\\
\hspace{1em}DL grading/survival  & 84.2(83.8,84.6) & 84.3(84.0,84.6) & 84.9(84.7,85.2) & 84.9(84.7,85.1) & 85.4(85.2,85.6) & 84.2(83.8,84.6)\\
Geographic atrophy &  &  &  &  &  & \\
\hspace{1em}Deep features/survival  & 89.9(89.6,90.1) & 90.1(89.8,90.4) & 88.5(88.2,88.8) & 88.5(88.2,88.7) & 89.0(88.8,89.2) & 88.7(88.5,88.9)\\
\hspace{1em}DL grading/survival  & 87.0(86.6,87.3) & 86.2(85.9,86.5) & 86.7(86.5,87.0) & 87.1(86.9,87.3) & 87.0(86.8,87.2) & 87.0(86.6,87.3)\\
Neovascular AMD &  &  &  &  &  & \\
\hspace{1em}Deep features/survival  & 83.7(83.2,84.2) & 77.6(77.0,78.3) & 81.5(81.0,81.9) & 81.9(81.6,82.3) & 81.3(80.9,81.6) & 82.3(82.1,82.6)\\
\hspace{1em}DL grading/survival  & 77.6(77.0,78.3) & 78.7(78.2,79.2) & 79.9(79.5,80.3) & 79.9(79.5,80.2) & 81.0(80.7,81.4) & 77.6(77.0,78.3)\\
\bottomrule
\end{tabular}
\end{center}
\end{table}

\subsection*{Research software prototype for AMD progression prediction}

\begin{figure}
	\centering
	\frame{\includegraphics[width=.75\textwidth,trim=0 14cm 0 2cm,clip]{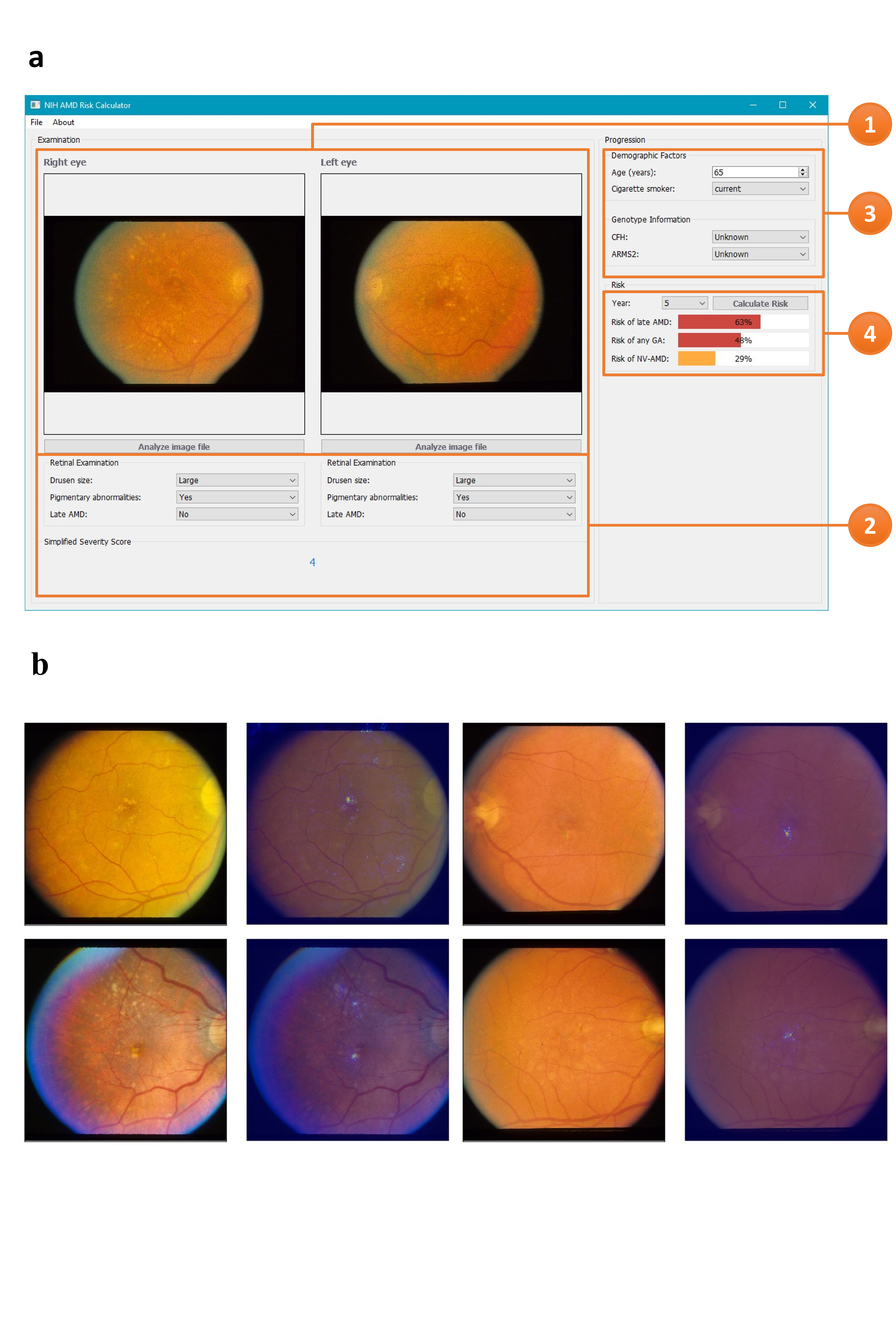}}
	\caption{\textbf{A screenshot of our research prototype system for AMD risk prediction.} (a) Screenshot of late AMD risk prediction. 1, Upload bilateral color fundus photographs. 2, Based on the uploaded images, the following information is automatically generated separately for each eye: drusen size status, pigmentary abnormality presence/absence, late AMD presence/absence, and the Simplified Severity Scale score. 3, Enter the demographic and (if available) genotype information, and the time point for prediction. 4, The probability of progression to late AMD (in either eye) is automatically calculated, along with separate probabilities of geographic atrophy and neovascular AMD. (b) Four selected color fundus photographs with highlighted areas used by the deep learning classification network (DeepSeeNet). Saliency maps were used to represent the visually dominant location (drusen or pigmentary changes) in the image by back-projecting the last layer of neural network.}
	\label{fig:screenshot}
\end{figure}

To demonstrate how these algorithms could be used in practice, we developed a software prototype that allows researchers to test our model with their own data. The application (shown in Figure~\ref{fig:screenshot}) receives bilateral CFP and performs autonomous AMD classification and risk prediction. For transparency, the researcher is given (i) grading of drusen and pigmentary abnormalities, (ii) predicted SSS, and (iii) estimated risks of late AMD, GA, and NV, over 1-12 years. This approach allows improved transparency and flexibility: users may inspect the automated gradings, manually adjust these if necessary, and recalculate the progression risks. Following further validation, this software tool may potentially augment human research and clinical practice.

\section*{Discussion}

We developed, trained, and validated a framework for predicting individual risk of late AMD by combining DL and survival analysis. This approach delivers autonomous predictions of a higher accuracy than those from retinal specialists using two existing clinical standards. Hence, the predictions are closer to the ground truth of actual time-based progression to late AMD than when retinal specialists are grading the same bilateral CFP and entering these grades into the SSS or the online calculator. In addition, deep feature extraction generally achieved slightly higher accuracy than DL grading of traditional hand-crafted features.

Table~\ref{tab:cindex5genotype} shows that the C-statistic values were lower for AREDS2, as expected, since the majority of its participants were at higher risk of progression\cite{ding2017bivariate}; though more difficult, predicting progression for AREDS2 participants is more representative of a clinically meaningful task. When compared to the results in Table~\ref{tab:cindex}, the accuracy of our models decreased less than retinal specialists/calculator and retinal specialists/SSS. This demonstrates the relative generalizability of our models. Furthermore, this suggests the ability of deep features/survival to differentiate individuals with relatively similar SSS more accurately than existing clinical standards (since nearly all AREDS2 participants had SSS $\leq$2 at baseline). It is also worth noting that it was possible to improve further the performance of our models on the AREDS2 dataset, if the models had been modified to accommodate the unique characteristics of the AREDS2. Since the AREDS is enriched for participants with milder baseline AMD severity, it may be possible to improve the performance of our models on the AREDS2 dataset by either training the models on progressors only or decreasing the prevalence of non-progressors. However, for fair comparisons to other results in this study, we kept our models unchanged. 

In addition, unlike the SSS, whose five-year risk prediction becomes saturated at 50\%10, the DL models enable ascertainment of risk above 50\%. This may be helpful in justifying medical and lifestyle interventions\cite{areds2001report8,areds2013lutein,guymer2019subthreshold}, vigilant home monitoring\cite{domalpally2019imaging,aredshomestudyresearchgroup2014randomized}, and frequent reimaging\cite{flaxel2020agerelateda}, and in planning shorter but highly powered clinical trials\cite{calaprice-whitty2019improving}. For example, the AREDS-style oral supplements decrease the risk of developing late AMD by approximately 25\%, but only in individuals at higher risk of disease progression\cite{areds2001report8,areds2013lutein}. Similarly, if subthreshold nanosecond laser treatment is approved to slow progression to late AMD8, accurate risk predictions may be very helpful for identifying eyes that may benefit most.

Considering that the DL grading model is better at detecting drusen and pigmentary changes, we also took the grades from the DL grading model and used these as input to the SSS and the Casey calculator, to generate predictions (Supplementary Table~\ref{tab:cindexcombine}). We found that the accuracy of the predictions was higher when using the grades from the DL grading than from the retinal specialists. Hence, it is the combination of the DL grading and the survival approach that provides accurate and automated predictions.

In addition, we compared C-statistic results on three DL models: nnet-survival\cite{gensheimer2019scalable}, DeepSurv\cite{katzman2018deepsurv}, and CoxPH (Supplementary Table~\ref{tab:cindexdeeplearning}). CoxPH had the best calibration performance using either the deep features or DL grading, though the differences were fairly small. 

For all approaches, the accuracy was substantially higher for predictions of GA than NV. The DL-based approaches improved accuracy for both subtypes, but, for NV, insufficiently to reach that obtained for GA. Potential explanations include the partially stochastic nature of NV and/or higher suitability of predicting GA from en face imaging. 

Error analysis (i.e., examining the reasons behind inaccurate predictions) was facilitated more easily by our two-step architecture, unlike in end-to-end, ‘black-box’ DL approaches. This revealed that the survival model accounted for most errors, since (i) classification network accuracy was relatively high (Supplementary Figure~\ref{fig:roc}), and (ii) when perfect classification results (i.e., taken from the ground truth) were used as input to the survival model, the five-year C-statistic of DL grading/survival on late AMD improved only slightly, from 85.1 (85.0,85.3) to 85.8 (85.6-86.0). Supplementary Figure~\ref{fig:examples} demonstrates two example cases of progression. In the first case, both participants were the same age (73 years) and had the same smoking history (current), and AREDS SSS scores (4) at baseline. The retinal specialist graded the SSS correctly. However, participant 1 progressed to late AMD at year 2 (any GA at right eye), but participant 2 progressed to late AMD at year 5 (any GA at left eye). Hence, both the retinal specialist/calculator and retinal specialist/SSS approaches incorrectly assigned the same risk of progression to both participants (0.413). However, the deep feature/survival approach correctly assigned a higher risk of progression to participant 1 (0.751) and a lower risk to participant 2 (0.591). In the second case, both participants were the same age (73 years) and had the same smoking history (former). Their AREDS SSS scores at baseline were 3 and 4, respectively. The retinal specialist graded the SSS correctly. However, participant 3 progressed to late AMD at year 2 (NV at right eye), while participant 4 had still not progressed to late AMD by final follow-up at year 11. Hence, both the retinal specialist/calculator and retinal specialist/SSS approaches incorrectly assigned a lower risk of progression to participant 3 (0.259) and a higher risk to participant 4 (0.413). However, the deep feature/survival approach correctly assigned a higher risk of progression to participant 4 and a lower risk to participant 3 (0.569 vs 0.528).

The strengths of the study include the application of combining DL image analysis and deep feature extraction with survival analysis to retinal disease. Survival analysis has been used widely in AMD progression research\cite{areds2001report8,ding2017bivariate}; in this study, it made best use of the data, specifically the timing and nature of any progression events at the individual level. Deep feature extraction has the advantages that the model is unconstrained by current medical knowledge and not limited to two features. It allows the model to learn de novo what features are most highly predictive of time-based progression events, and to develop multiple (e.g., 5, 10, or more) predictive features. Unlike shallow features that appear early in CNNs, deep features are potentially complex and high-order features, and might even be relatively invisible to the human eye. However, one limitation of deep feature extraction is that predictions based on these features may be less explainable and biologically plausible, and less amenable to error analysis, than those based on DL grading of traditional risk features\cite{ting2019deep}. Additional strengths include the use of two well-characterized cohorts, with detailed time-based knowledge of progression events at the reading center standard. By pooling AREDS and AREDS2 (which has not been used previously in DL studies), we were able to construct a cohort that had a wide spectrum of AMD severity but was enriched for cases of higher baseline severity. In addition, in other experiments, keeping the datasets separate enabled us to perform external validation using AREDS2 as an independent cohort.

In terms of limitations, as in the two existing clinical standards, AREDS/AREDS2 treatment assignment was not considered in this analysis10,12. Since most AREDS/AREDS2 participants were assigned to oral supplements that decreased risk of late AMD, the risk estimates obtained are closer to those for individuals receiving supplements. However, this seems appropriate, given that AREDS-style supplements are considered the standard of care for patients with intermediate AMD\cite{flaxel2020agerelateda}. Another limitation is that this work relates to CFP only. DL approaches to optical coherence tomography (OCT) datasets hold promise for AMD diagnosis\cite{defauw2018clinically,lee2017deep}, but no highly validated OCT-based tools exist for risk prediction. In the future, we plan to apply our framework to multi-modal imaging, including fundus autofluorescence and OCT data.

In conclusion, combining DL feature extraction of CFP with survival analysis achieved high prognostic accuracy in predictions of progression to late AMD, and its subtypes, over a wide time interval (1-12 years). Not only did its accuracy meet and surpass existing clinical standards, but additional strengths in clinical settings include risk ascertainment above 50\% and without genotype data.

\newpage

\section*{Methods}

\subsection*{Datasets}

For model development and clinical validation, two datasets were used: the AREDS\cite{areds1999report1} and the AREDS2\cite{areds22012report1} (Figure~\ref{fig:data}). The AREDS was a 12-year multi-center prospective cohort study of the clinical course, prognosis, and risk factors of AMD, as well as a phase III randomized clinical trial (RCT) to assess the effects of nutritional supplements on AMD progression\cite{areds1999report1}. In short, 4,757 participants aged 55 to 80 years were recruited between 1992 and 1998 at 11 retinal specialty clinics in the United States. The inclusion criteria were wide, from no AMD in either eye to late AMD in one eye. The participants were randomly assigned to placebo, antioxidants, zinc, or the combination of antioxidants and zinc. The AREDS dataset is publicly accessible to researchers by request at dbGAP (\url{https://www.ncbi.nlm.nih.gov/projects/gap/cgi-bin/study.cgi?study_id=phs000001.v3.p1}).

Similarly, the AREDS2 was a multi-center phase III RCT that analyzed the effects of different nutritional supplements on the course of AMD32. 4,203 participants aged 50 to 85 years were recruited between 2006 and 2008 at 82 retinal specialty clinics in the United States. The inclusion criteria were the presence of either bilateral large drusen or late AMD in one eye and large drusen in the fellow eye. The participants were randomly assigned to placebo, lutein/zeaxanthin, docosahexaenoic acid (DHA) plus eicosapentaenoic acid (EPA), or the combination of lutein/zeaxanthin and DHA plus EPA. AREDS supplements were also administered to all AREDS2 participants, because they were by then considered the standard of care\cite{flaxel2020agerelateda}. We will make the AREDS2 dataset publicly accessible upon publication. 

In both studies, the primary outcome measure was the development of late AMD, defined as neovascular AMD or central GA. Institutional review board approval was obtained at each clinical site and written informed consent for the research was obtained from all study participants. The research was conducted under the Declaration of Helsinki and, for the AREDS2, complied with the Health Insurance Portability and Accessibility Act. For both studies, at baseline and annual study visits, comprehensive eye examinations were performed by certified study personnel using a standardized protocol, and CFP (field 2, i.e., 30\textdegree{} imaging field centered at the fovea) were captured by certified technicians using a standardized imaging protocol. Progression to late AMD was defined by the study protocol based on the grading of CFP\cite{areds1999report1,areds22012report1}, as described below.

As part of the studies, 2,889 (AREDS) and 1,826 (AREDS2) participants consented to genotype analysis. SNPs were analyzed using a custom Illumina HumanCoreExome array\cite{fritsche2016large}. For the current analysis, two SNPs (CFH rs1061170 and ARMS2 rs10490924, at the two loci with the highest attributable risk of late AMD), were selected, as these are the two SNPs available as input for the existing online calculator system. In addition, the AMD GRS was calculated for each participant according to methods described previously\cite{fritsche2016large}. The GRS is a weighted risk score based on 52 independent variants at 34 loci identified in a large genome-wide association study\cite{fritsche2016large} as having significant associations with risk of late AMD. The online calculator cannot receive this detailed information.

The eligibility criteria for participant inclusion in the current analysis were: (i) absence of late AMD (defined as NV or any GA) at study baseline in either eye, since the predictions were made at the participant level, and (ii) presence of genetic information (in order to compare model performance with and without genetic information on exactly the same cohort of participants). Accordingly, the images used for the predictions were those from the study baselines only.

In the AREDS dataset of CFPs, information on image laterality (i.e., left or right eye) and field status (field 1, 2, or 3) were available from the Reading Center. However, these were not available in the AREDS2 dataset of CFPs. We therefore trained two Inception-v3 models, one for classifying laterality and the other for identifying field 2 images. Both models were first trained on the gold standard images from the AREDS and fine-tuned on a newly created gold standard AREDS2 set manually graded by a retinal specialist (TK). The AREDS2 gold standard consisted of 40 participants with 5,164 images (4,097 for training and 1,067 for validation). The models achieved 100\% accuracy for laterality classification and 97.9\% accuracy (F1-score 0.971, precision 0.968, recall 0.973) for field 2 classification.

\subsection*{Gold standard grading}

The ground truth labels used for both training and testing were the grades previously assigned to each CFP by expert human graders at the University of Wisconsin Fundus Photograph Reading Center. The reading center workflow has been described previously\cite{areds2001report6}. In brief, a senior grader performed initial grading of each photograph for AMD severity using a 4-step scale and a junior grader performed detailed grading for multiple AMD-specific features. All photographs were graded independently and without access to the clinical information. A rigorous process of grading quality control was performed at the reading center, including assessment for inter-grader and intra-grader agreement\cite{areds2001report6}. The reading center grading features relevant to the current study, aside from late AMD, were: (i) macular drusen status (none/small, medium (diameter $\leq$63µm and $<$125µm), and large ($\leq$125µm)), and (ii) macular pigmentary abnormalities related to AMD (present or absent).

In addition to undergoing reading center grading, the images at the study baseline were also assessed (separately and independently) by 88 retinal specialists in AREDS and 196 retinal specialists in AREDS2. The responses of the retinal specialists were used not as the ground truth, but for comparisons between human grading as performed in routine clinical practice and DL-based grading. By applying these retinal specialist grades as input to the two existing clinical standards for predicting progression to late AMD, it was possible to compare the current clinical standard of human predictions with those predictions achievable by DL.

\subsection*{Development of the algorithm}

The overall framework of our method is shown in Figure~\ref{fig:architecture}. First, a CNN was adapted to (i) extract multiple highly discriminative deep features, or (ii) estimate grades for drusen and pigmentary abnormalities (Figure~\ref{fig:architecture} a to c). Second, a Cox proportional hazards model was used to predict probability of progression to late AMD (and GA/NV, separately), based on the deep features (‘deep features/survival’) or the DL grading (‘DL grading/survival’) (Figure~\ref{fig:architecture} e and e). In this step, additional participant information could be added, such as age, smoking status, and genetics.

As the first stage in the workflow, the DL-based image analysis was performed using two different adaptations of ‘DeepSeeNet’\cite{peng2018deepseenet}. DeepSeeNet is a CNN framework that was created for AMD severity classification. It has achieved state-of-the-art performance for the automated diagnosis and classification of AMD severity from CFP; this includes the grading of macular drusen, pigmentary abnormalities, the SSS\cite{peng2018deepseenet}, and the AREDS 9-step severity scale\cite{chen2019multitaska}. In particular, using reading center grades as the ground truth, we have recently demonstrated that DeepSeeNet performs grading with accuracy that was superior to that of human retinal specialists (Supplementary Figure~\ref{fig:survival}). The two different adaptations are described here:

\textbf{Deep features}. The first adaptation was named ‘deep features’. This approach involved using DL to derive and weight predictive image features, including high-dimensional ‘hidden’ features\cite{lao2017deep}. Deep features were extracted from the second to last fully-connected layer of DeepSeeNet (the highlighted part in the classification network in Figure~\ref{fig:architecture}). In total, 512 deep features could be extracted for each participant in this way, comprising 128 deep features for each of the two models (drusen and pigmentary abnormalities) in each of the two images (left and right eyes). After feature extraction, all 512 deep features were normalized as standard-scores. Feature selection was required at this point, to avoid overfitting and to improve the generalizability, because of the multi-dimensional nature of the features. Hence, we performed feature selection to group correlated features and pick one feature for each group\cite{simon2011regularization}. Features with non-zero coefficients were selected and applied as input to the survival models described below.

\textbf{Deep learning grading.} The second adaptation of DeepSeeNet was named ‘DL grading’, i.e., referring to the grading of drusen and pigmentary abnormalities, the two macular features considered by humans most able to predict progression to late AMD. In this adaptation, the two predicted risk factors were used directly. In brief, one CNN was previously trained and validated to estimate drusen status in a single CFP, according to three levels (none/small, medium, or large), using reading center grades as the ground truth\cite{peng2018deepseenet}. A second CNN was previously trained and validated to predict the presence or absence of pigmentary abnormalities in a single CFP.

\textbf{Survival model}. The second stage of our workflow comprised a Cox proportional hazards model\cite{cox1992regression} to estimate time to late AMD (Fig1. D and E). The Cox model is used to evaluate simultaneously the effect of several factors on the probability of the event, i.e., participant progression to late AMD in either eye. Separate Cox proportional hazards models were created to analyze time to late AMD and time to subtype of late AMD (i.e., GA and NV). In addition to the image-based information, the survival models could receive three additional inputs: (i) participant age; (ii) smoking status (current/former/never), and (iii) participant AMD genotype (CFH rs1061170, ARMS2 rs10490924, and the AMD GRS).

\subsection*{Experimental design}

In both of the DeepSeeNet adaptations described, the DL CNNs used Inception-v3 architecture\cite{szegedy2016rethinking}, which is a state-of-the-art CNN for image classification; it contains 317 layers, comprising a total of over 21 million weights that are subject to training. Training was performed using two commonly used libraries: Keras (https://keras.io) and TensorFlow\cite{abadi2016tensorflow}. All images were cropped to generate a square image field encompassing the macula and resized to 512 x 512 pixels. The hyperparameters were learning rate 0.0001 and batch size 32. The training was stopped after 5 epochs once the accuracy on the development set no longer increased. All experiments were conducted on a server with 32 Intel Xeon CPUs, using a NVIDIA GeForce GTX 1080 Ti 11Gb GPU for training and testing, with 512Gb available in RAM memory. We fitted the Cox proportional hazard model using the deep features as covariates. Specifically, we selected the 16 features with the highest weights (i.e. the features found to be most predictive of progression to late AMD, with their inclusion as covariates in the Cox model). We performed feature selection using the ‘glmnet’ package\cite{simon2011regularization} in R version 3.5.2 statistical software.

\subsection*{Training and testing}

For training and testing our framework, we used both the AREDS and AREDS2 datasets. In the primary set of experiments, eligible participants from both studies were pooled to create one broad cohort of 3,298 individuals that combined a wide spectrum of baseline disease severity with a high number of progression events. The combined dataset was split at the participant level in the ratio 70\%/10\%/20\% to create three sets: 2,364 participants (training set), 333 participants (development set), and 601 participants (hold-out test set).

Separately, all of the baseline images in the test set were graded by 88 (AREDS) and 192 (AREDS2) retinal specialists. By using these grades as input to either the SSS or the online calculator, we computed the prediction results of the two existing standards: ‘retinal specialists/SSS’ and ‘retinal specialists/calculator’.

For three of the four approaches (deep features/survival, DL grading/survival, and retinal specialists/calculator), the input was bilateral CFP, participant age, and smoking status; separate experiments were conducted with and without the additional input of genotype data. For the other approach (retinal specialists/SSS), the input was bilateral CFP only.

In addition to the primary set of experiments where eligible participants from the AREDS and AREDS2 were combined to form one dataset, separate experiments were conducted where the DL models were: (i) trained separately on the AREDS training set only, or the AREDS2 training set only, and tested on the combined AREDS/AREDS2 test set, and (ii) trained on the AREDS training set only and externally validated by testing on the AREDS2 test set only.

\subsection*{Statistical analysis}

As the primary outcome measure, the performance of the risk prediction models was assessed by the C-statistic\cite{pencina2004overall} at five years from study baseline. Five years from study baseline was chosen as the interval for the primary outcome measure since this is the only interval where comparison can be made with the SSS, and the longest interval where predictions can be tested using the AREDS2 data.

For binary outcomes such as progression to late AMD, the C-statistic represents the area under the receiver operating characteristic curve (AUC). The C-statistic is computed as follows: all possible pairs of participants are considered where one participant progressed to late AMD and the other participant in the pair progressed later or not at all; out of all these pairs, the C-statistic represents the proportion of pairs where the participant who had been assigned the higher risk score was the one who did progress or progressed earlier. A C-statistic of 0.5 indicates random predictions, while 1.0 indicates perfectly accurate predictions. We used 200 bootstrap samples to obtain a distribution of the C-statistic and reported 95\% confidence intervals. For each bootstrap iteration, we sampled n patients with replacement from the test set of n patients.

As a secondary outcome measure of performance, we calculated the Brier score from prediction error curves, following the work of Klein et al\cite{klein2011risk}. The Brier score is defined as the squared distances between the model’s predicted probability and actual late AMD, GA, or NV status, where a score of 0.0 indicates a perfect match. The Wald test was used to assess the statistical significance of each factor in the survival models\cite{rosner2015fundamentals}. It corresponds to the ratio of each regression coefficient to its standard error. The ‘survival’ package in R version 3.5.2 was used for Cox proportional hazards model evaluation. Finally, saliency maps were generated to represent the image locations that contributed most to decision-making by the DL models (for drusen or pigmentary abnormalities). This was done by back-projecting the last layer of the neural network. The Python package ‘keras-vis’ was used to generate the saliency map\cite{simonyan2013deep}.

\section*{Acknowledgements}
The work was supported by the intramural program funds and contracts from the National Center for Biotechnology Information/National Library of Medicine/National Institutes of Health, the National Eye Institute/National Institutes of Health, Department of Health and Human Services, Bethesda Maryland (Contract HHS-N-260-2005-00007-C; ADB contract NO1-EY-5-0007; Grant No 4R00LM013001). Funds were generously contributed to these contracts by the following National Institutes of Health: Office of Dietary Supplements, National Center for Complementary and Alternative Medicine; National Institute on Aging; National Heart, Lung, and Blood Institute; and National Institute of Neurological Disorders and Stroke.

\bibliographystyle{vancouver}
\bibliography{references}

\pagebreak

\section*{Supplementary information}
\setcounter{figure}{0}
\setcounter{table}{0}
\renewcommand{\thefigure}{S\arabic{figure}}
\renewcommand{\thetable}{S\arabic{table}}

\begin{figure}[H]
	\centering
	\frame{\includegraphics[width=.8\textwidth]{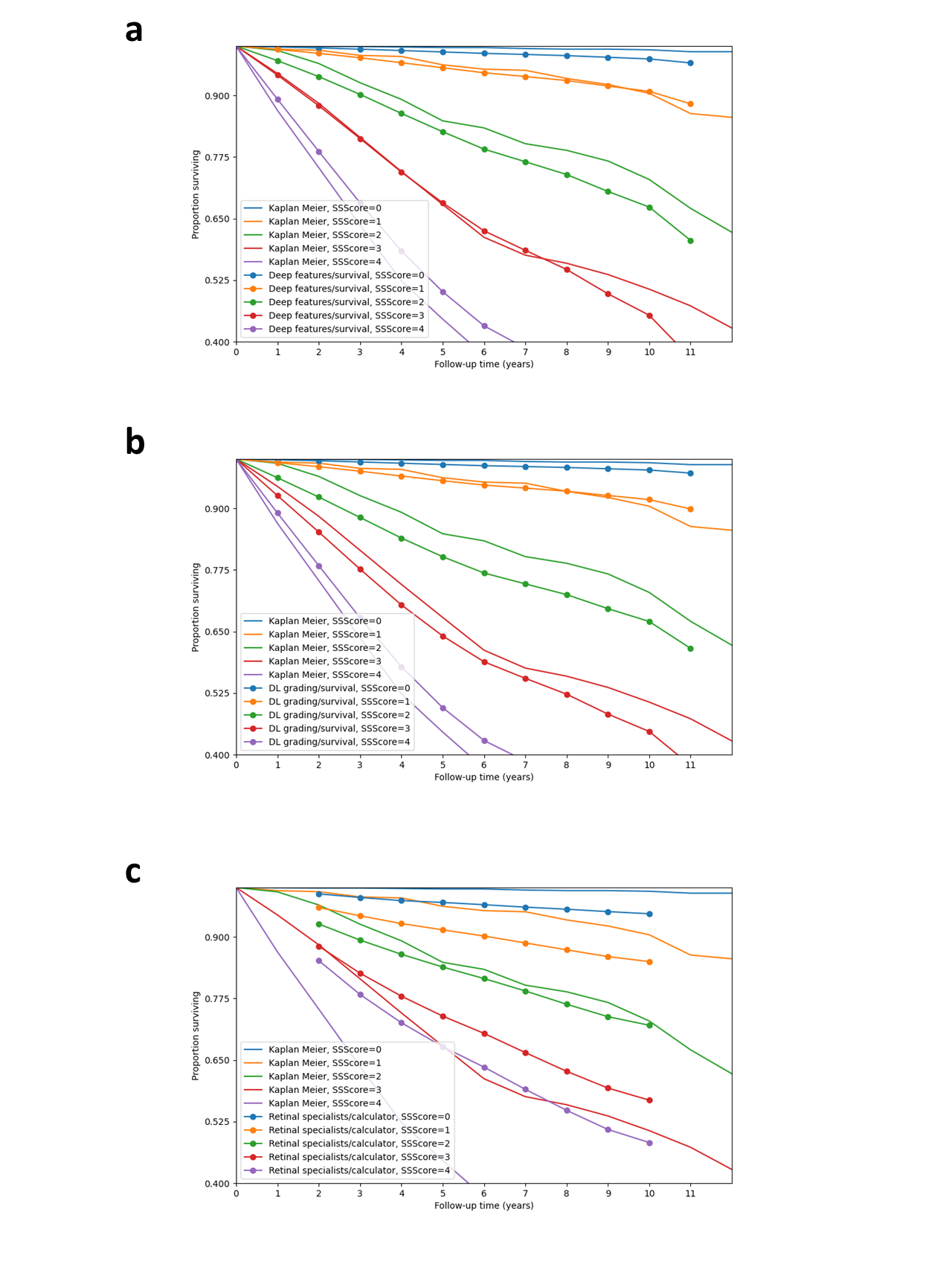}}
	\caption{\textbf{Survival model predictions for the 5 groups.} The participants were split into 5 groups based on the AREDS simplified severity scores at the baseline. Actual survival for the 5 groups is shown in lines (Kaplan-Meier curves). The deep features/survival, DL grading/survival, and retinal specialist/survival model predictions for the five groups are shown in lines with markers. The deep features/survival predictions corresponded better to actual progression data than those of the other two models.}
	\label{fig:survival}
\end{figure}

\pagebreak

\begin{figure}[H]
	\centering
	\frame{\includegraphics[width=\textwidth,trim=0 4cm 0 3cm,clip]{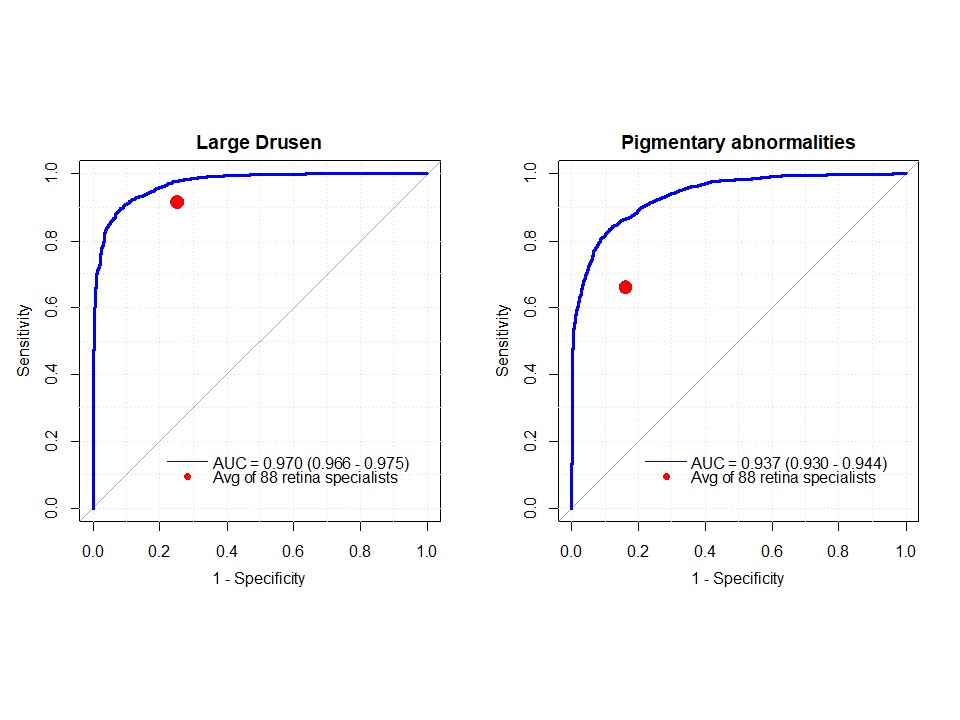}}
	\caption{\textbf{Receiver operating characteristic curves for the deep neural networks to grade drusen size and pigmentary abnormalities presence/absence on the combined AREDS/AREDS2 test sets (601 participants).} The AUC was used to evaluate the performance of deep neural networks in image classification (as opposed to predictions of progression), and for comparison with the retinal specialists (with reference to the Reading Center grades as the ground truth).}
	\label{fig:roc}
\end{figure}

\pagebreak

\begin{figure}[H]
	\centering
	\frame{\includegraphics[width=.75\textwidth,trim=0 2cm 0 0,clip]{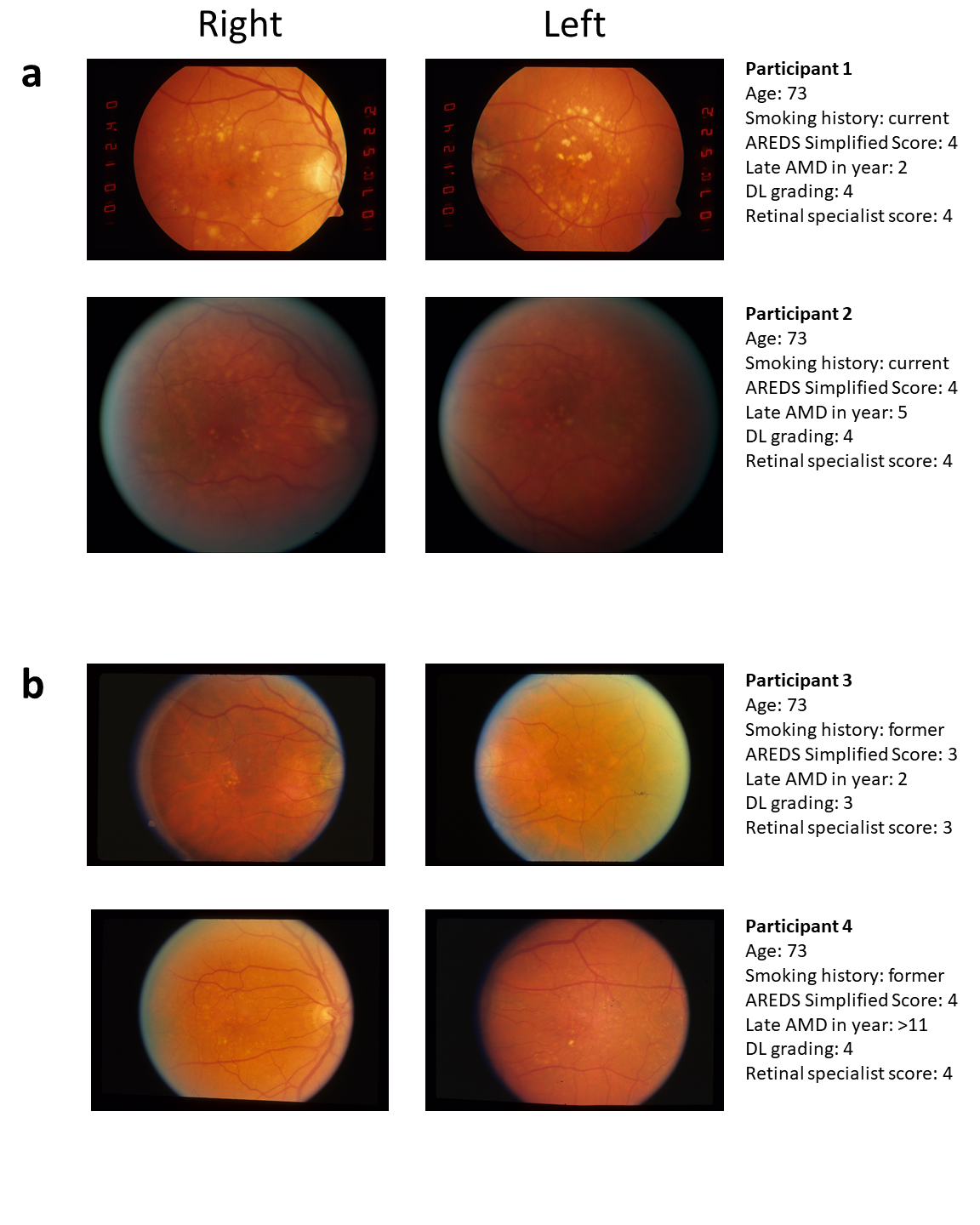}}
	\caption{\textbf{Example cases of progression to late age-related macular degeneration (AMD) where deep features/survival made more accurate predictions than retinal specialist/calculator or retinal specialist/Simplified Severity Scale (SSS).} a. Both participants were the same age (73 years) and had the same smoking history (current) and AREDS SSS scores (4) at baseline. The retinal specialist graded the SSS correctly. However, participant 1 progressed to late AMD at year 2 (GA in the right eye), but participant 2 progressed to late AMD at year 5 (GA in the left eye). Hence, both the retinal specialist/calculator and retinal specialist/SSS approaches \textit{\underline{incorrectly}} assigned the same risk of progression to both participants (0.413). However, the deep feature/survival approach correctly assigned a higher risk of progression to participant 1 (0.751) and a lower risk to participant 2 (0.591). b. Both participants were the same age (73 years) and had the same smoking history (former). Their AREDS SSS scores at baseline were 3 and 4, respectively. The retinal specialist graded the SSS correctly. However, participant 3 progressed to late AMD at year 2 (neovascular AMD in the right eye), while participant 4 had still not progressed to late AMD by final follow-up at year 11. Hence, both the retinal specialist/calculator and retinal specialist/SSS approaches \textit{\underline{incorrectly}} assigned a lower risk of progression to participant 3 (0.259) and a higher risk to participant 4 (0.413). However, the deep feature/survival approach \textit{\underline{correctly}} assigned a higher risk of progression to participant 4 and a lower risk to participant 3 (0.569 vs 0.528).}
	\label{fig:examples}
\end{figure}

\pagebreak

\begin{table}[!ht]
\caption{Number of images used to develop the classification network.}
\label{tab:number}
\begin{center}
\begin{tabular}{lrr}
 \toprule 
Image characteristics & AREDS & AREDS2\\
\midrule
Number of images 										& 57,375 & 17,658\\
\hspace{1em}Training 								& 40,455 & 12,391\\
\hspace{1em}Development 							&  5,535 & 1,738\\
\hspace{1em}Test 										& 11,385 & 3,529\\
\multicolumn{2}{l}{AMD drusen feature, by image, as classified by Reading Center}\\
\hspace{1em}No drusen or small drusen, No. (\%) 	& 24,290 & 191\\
\hspace{1em}Medium drusen, No. (\%) 				& 15,464 & 921\\
\hspace{1em}Large drusen, No. (\%) 					& 17,621 & 16,546 \\
\multicolumn{2}{l}{AMD pigmentary abnormalities, by image, as classified by Reading Center}\\
\hspace{1em}absent, No. (\%) 							& 40,738 & 3,855\\
\hspace{1em}present, No. (\%) 						& 16,637 & 13,803\\
 \bottomrule
\end{tabular}
\end{center}
\end{table}

\pagebreak

\begin{table}[!ht]
\caption{Multivariate Association of Phenotypic, Demographic, and Genetic Variables and Progression to late AMD. Deep features with p value $<0.05$ are shown.}
\label{tab:multivariate}
\begin{center}
\small
\begin{tabular}{lccc|lccc}
\toprule 
\multicolumn{4}{c|}{DL grading/survival} & \multicolumn{4}{c}{Deep Feature/survival}\\
Variable & Hazard  & 95\% CI & p-value & Variable & Hazard  & 95\% CI & p-value\\
         & ratio	 &         &         &          & ratio   &         &\\
\midrule
Age 					& 1.05 & 1.04-1.07 & $<$.001 	& Age 				& 1.05 & 1.04-1.06 & $<$.001\\
Smoking status 	& 1.17 & 1.03-1.32 & 0.017 	& Smoking status 	& 1.14 & 1.00-1.30 & 0.055\\
CFH rs1061170 		& 0.95 & 0.83-1.07 & 0.382 	& CFH rs1061170 	& 0.89 & 0.77-1.04 & 0.085\\
ARMS2 rs10490924 	& 0.95 & 0.83-1.09 & 0.490 	& ARMS2 rs10490924& 0.90 & 0.77-1.04 & 0.161\\
AMD GRS 				& 1.35 & 1.24-1.47 & $<$.001	& AMD GRS 			& 1.38 & 1.25-1.52 & $<$.001\\
drusen score (LE) & 2.26 & 1.84-2.78 & $<$.001 	& feature335		& 0.14 & 0.05-0.37 & $<$.001\\
drusen score (RE) & 1.89 & 1.56-2.28 & $<$.001 	& feature79 		& 0.24 & 0.09-0.66 & 0.005\\
pig abn (LE) 		& 2.28 & 1.88-2.76 & $<$.001 	& feature449 		& 0.56 & 0.35-0.91 & 0.020\\
pig abn (RE) 		& 1.72 & 1.42-2.08 & $<$.001 	& feature234 		& 9.41 & 1.04-85.11& 0.046\\
\bottomrule
\end{tabular}
\end{center}
LE-left eye, RE-right eye, pig abn-pigmentary abnormality, GRS-Genetic Risk Score.
\end{table}

\pagebreak

\begin{table}[!ht]
\caption{ The C-statistic (95\% confidence interval) of the survival models in predicting risk of progression to late age-related macular degeneration on the combined AREDS/AREDS2 test sets (601 participants).}
\label{tab:cindexcombine}
\begin{center}
\footnotesize
\begin{tabular}{l>{\centering}m{2cm}cccc}
\toprule 
Models & 1 & 2 & 3 & 4 & 5\\
\midrule
Late AMD 									&   &  &  &  & \\
\hspace{1em}DL grading/calculator 	& - & 82.2 (81.8, 82.6) & 82.6 (82.3, 83.0) & 83.4 (83.1, 83.6) & 83.5 (83.3, 83.7)\\
\hspace{1em}DL grading/SSS* 			& - & - & - & - & 82.3 (82.1, 82.5)\\
Geographic atrophy 						&   &  &  &  & \\
\hspace{1em}DL grading/calculator 	& - & 84.7 (84.4, 85.1) & 83.4 (83.1, 83.7) & 84.2 (83.9, 84.4) & 83.7 (83.4, 83.9)\\
\hspace{1em}DL grading/SSS* 			& - & - & - & - & -\\
Neovascular AMD 							&   &  &  &  & \\
\hspace{1em}DL grading/calculator 	& - & 78.2 (77.4, 79.0) & 80.1 (79.5, 80.6) & 78.7 (78.3, 79.1) & 79.4 (79.1, 79.7)\\
\hspace{1em}DL grading/SSS* 			& - & - & - & - & -\\
\bottomrule
\end{tabular}
\end{center}
\end{table}

\pagebreak
\begin{table}[!ht]
\caption{The C-statistic (95\% confidence interval) of three models in predicting risk of progression to late age-related macular degeneration on the combined AREDS/AREDS2 test sets (601 participants).}
\label{tab:cindexdeeplearning}
\begin{center}
\begin{tabular}{lcc}
\toprule 
               & DL grading 		   & DL features\\
\midrule
Nnet-survival 	& 0.803 (0.799, 0.808) & 0.851 (0.846, 0.856)\\
DeepSurv 		& 0.842 (0.837, 0.847) & 0.859 (0.854, 0.865)\\
COX PH 			& 0.849 (0.846, 0.853) & 0.867 (0.865, 0.868)\\
\bottomrule
\end{tabular}
\end{center}
\end{table}

\end{document}